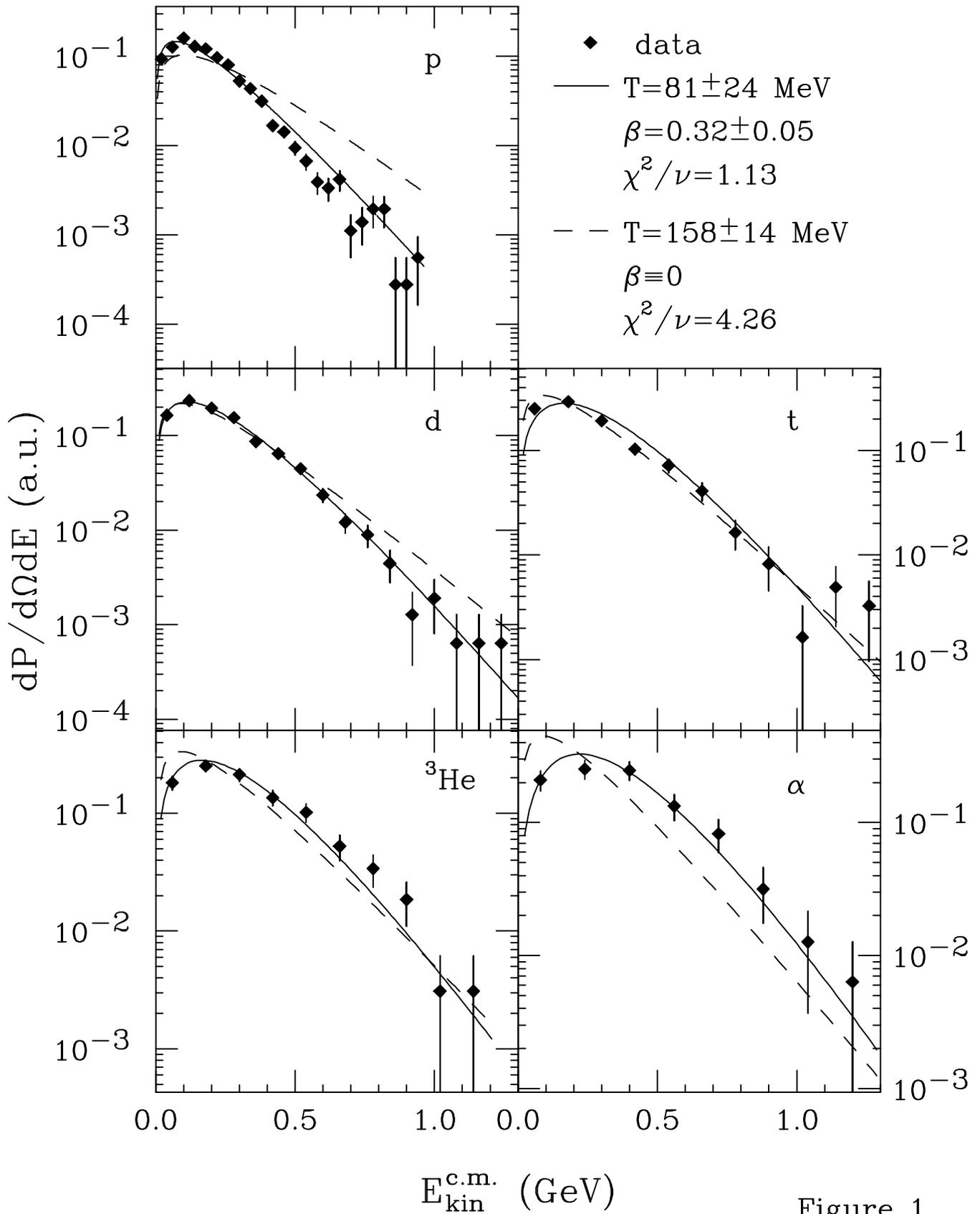

Figure 1

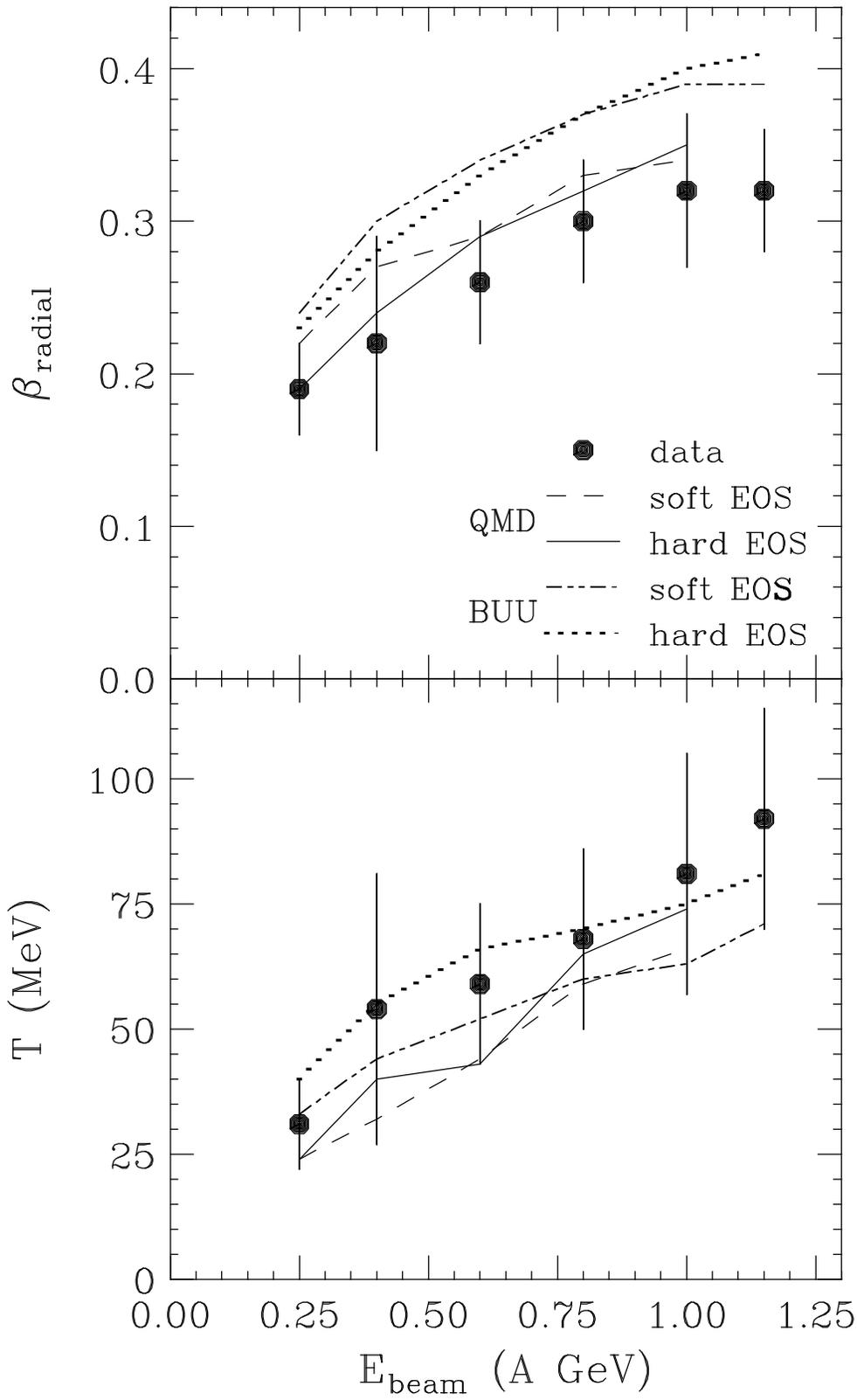

Figure 2

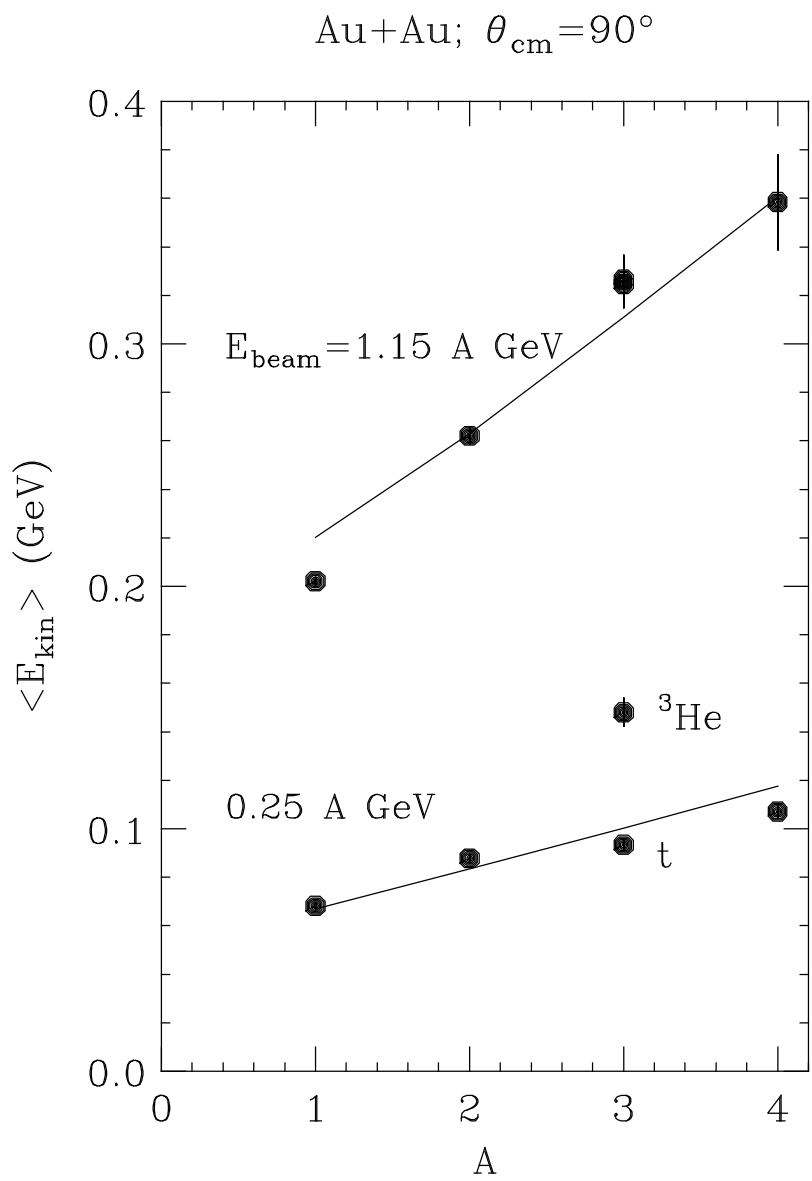

Figure 3

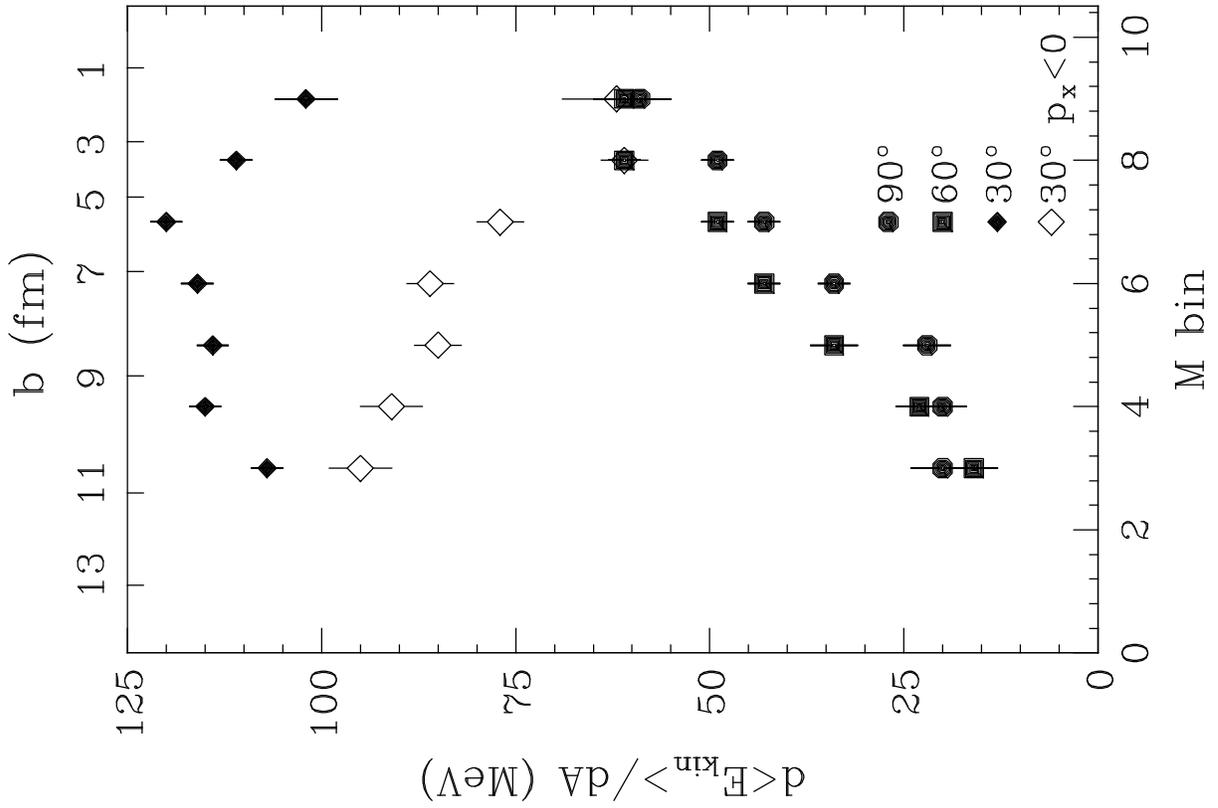

Figure 4

# Radial Flow in Au+Au Collisions at E=0.25-1.15 A GeV

M.A. Lisa[1], S. Albergo[6], F. Bieser[1], F. P. Brady[4], Z. Caccia[6], D. A. Cebra[4], A.D. Chacon[5], J. L. Chance[4], Y. Choi[3*], S. Costa[6], J. B. Elliott[3], M. L. Gilkes[3†], J. A. Hauger[3], A. S. Hirsch[3], E. L. Hjort[3], A. Insolia[6], M. Justice[2], D. Keane[2], J. Kintner[4], H. S. Matis[1], M. McMahan[1], C. McParland[1], D.L. Olson[1], M. D. Partlan[4], N. T. Porile[3], R. Potenza[6], G. Rai[1], J. Rasmussen[1], H.G. Ritter[1], J. Romanski[6], J. L. Romero[4], G. V. Russo[6], R. Scharenberg[3], A. Scott[2], Y. Shao[2‡], B. K. Srivastava[3], T.J.M. Symons[1], M. Tincknell[3], C. Tuvé[6], S. Wang[2], P. Warren[3], G.D. Westfall[1,7], H.H. Wieman[1], K. Wolf[5]

[1] Nuclear Science Division, Lawrence Berkeley Laboratory, Berkeley, CA, 94720
[2] Kent State University, Kent, Ohio 44242
[3] Purdue University, West Lafayette, Indiana, 47907-1396
[4] University of California, Davis, California, 95616
[5] Texas A&M University, College Station, Texas, 77843
[6] Università di Catania & INFN-Sezione di Catania, Catania, Italy, 95129
[7] NSCL, Michigan State University, East Lansing, MI 48824

[*] Current address: Sung Kwun Kwan University, Suwon 440-746, Republic of Korea
[†] Current address: Depts. of Physics and Chemistry, SUNY at Stony Brook, Stony Brook, NY 11794
[‡] Current address: Department of Pharmacology, UCLA School of Medicine, Los Angeles, CA 91776

The EOS Collaboration

**Abstract**: A systematic study of energy spectra for light particles emitted at midrapidity from Au+Au collisions at E=0.25-1.15 A GeV reveals a significant non-thermal component consistent with a collective radial flow. This component is evaluated as a function of bombarding energy and event centrality. Comparisons to Quantum Molecular Dynamics (QMD) and Boltzmann-Uehling-Uhlenbeck (BUU) models are made for different equations of state.

PACS numbers: 25.75.+r, 25.70.Gh

Collective motion plays an important role in the decay of excited nuclear matter and has been studied over a wide range of bombarding energies in heavy ion collisions [1-10]. Much of the initial interest in collective motion centered on the possibility that the nuclear equation of state may be extracted from its measurement [11]. Directed collective flow is experimentally well-established and has been used to study the nuclear equation of state [12]. However, only a small fraction of the energy available in the center of momentum is contained in directed flow, whereas from entropy studies [13] and general energy considerations [14], one expects a large portion of the total energy to be contained in collective degrees of freedom. With the recent availability of high-statistics exclusive measurements, there are indications that a large radial component of collective flow exists in central heavy ion collisions at beam energies around 100 A MeV [9, 10]. The EOS collaboration has recently completed a systematic measurement of heavy ion collisions at Bevalac energies, providing the opportunity to study radial flow as a function of bombarding energy and impact parameter. We observe a strong signal for radial flow in Au+Au collisions by studying spectral shapes for particles emitted at midrapidity. Model calculations agree well with the data and indicate little sensitivity of the radial flow to the nuclear equation of state.

The data were taken at Lawrence Berkeley Laboratory using Au beams from the Bevalac with bombarding energies in the range E = 0.25-1.15 A GeV incident on a gold target. Reaction products were measured by the detector systems that comprised the EOS experimental setup. This setup included a Time Projection Chamber (TPC) [15], a multi-sampling ion chamber [16], a time-of-flight wall, a neutron spectrometer [17], and beam diagnostic detectors. This paper is concerned only with data measured in the TPC.

The TPC is well suited to search for radial flow. It has good particle identification, especially for the hydrogen and helium isotopes [18]. An measurement of the radial flow is best obtained by studying several particle species with different masses, since a particle's energy due to flow is proportional to the mass, while its thermal energy is not [1-6, 9]. The TPC has no low-$p_T$ detection threshold, and the data were taken with high statistics, allowing a careful analysis of the shape of the energy spectra. The TPC provides excellent acceptance at midrapidity $\theta_{cm} \approx 90°$. Searching for radial flow exclusively at other angles introduces the risk of contaminating the signal with directed flow effects and emission from non-participant sources. Consequently, studies of spectra away from midrapidity are forced to employ very stringent centrality cuts that attempt to select spherically symmetric events [9]. By concentrating on particles emitted at $\theta_{cm}=90°\pm15°$, we are able to focus on the participant source and also can explore the impact parameter dependence of the radial flow of the participant source, which may be important in order to fully understand the effect. In our analysis, the event centrality is determined by multiplicity cuts similar to those used by the Plastic Ball group [19]. $M_{max}$ is defined as the multiplicity at which the multiplicity distribution assumes a value of half the plateau value. The region M=0-$M_{max}$ is divided into 8 equal-width bins. The most central events have M>$M_{max}$ and fall into bin 9.

The energy distribution in the center of mass for particles emitted from a thermally-equilibrated, radially-expanding source, characterized by a temperature T and a radial flow velocity $\beta$, is given by the functional form [1]:

$$\frac{d^3N}{dEd^2\Omega} \sim p \cdot e^{-\gamma E/T} \left\{ \frac{\sinh \alpha}{\alpha} \cdot (\gamma E + T) - T \cdot \cosh \alpha \right\}, \qquad (1)$$

where E and p are the total energy and momentum of the particle in the center of mass, $\gamma=(1-\beta^2)^{-1/2}$, and $\alpha = \gamma\beta p/T$. (Equation (1) differs from that in Ref. [1] by the relativistic Jacobian factor E·p.) Although somewhat schematic, the concept of a source provides a useful way to parameterize the data and identify important components in the decay of the excited system.

We have extracted source temperature and radial flow velocity parameters by fitting the functional form (1) to the energy spectra measured at $\theta_{cm}=90°\pm15°$, using a $\chi^2/\nu$ (chi-squared per degree of freedom) minimization technique, and assuming statistical errors for the data points. In Figure 1, we show kinetic energy spectra, measured at $\theta_{cm}=90°\pm15°$, for hydrogen and helium isotopes for the reaction Au+Au at E=1.0 A GeV. The data are from the most central (highest multiplicity) events, corresponding to less than 5% of the total cross-section (b≈0-3 fm in a geometrical picture). Also shown are fits to the spectrum with the form (1). Solid lines indicate a simultaneous fit to all spectra, excluding the proton spectrum (see below), by varying $\beta$ and T, and fixing the relative normalization of the fits for different particle types to match measured relative yields. A good overall fit is obtained, with a $\chi^2/\nu$ on the order of unity. Stated uncertainties are the 1-$\sigma$ errors of the fit. Dashed lines show fits with a purely thermal scenario ($\beta=0$). The spectral shapes are not as well reproduced, especially for the heavier fragments. At all bombarding energies considered, fits with nonzero flow consistently yield $\chi^2/\nu$ 2-4 times smaller than thermal fits. Fits to event-generated spectra before and after filtering through the simulated detector response indicate that temperature and flow velocity change by less than 10% and 4% respectively, well within the statistical uncertainties.

When the spectra for the fragment types (d,t,$^3$He,$\alpha$) are fit separately, the extracted temperature and flow values are consistent with the values obtained with the simultaneous fit of all particle types. However, fit parameters for proton spectra consistently indicate a lower temperature (by ~20%) and greater flow (by about 0.06c). Calculations with a fireball model [20] indicate that these deviations can be qualitatively understood in terms of distortions of the proton spectrum due to baryonic (e.g. $\Delta$) and nuclear (e.g. $^5$Li) resonance decay.

Coulomb repulsion from the emitting system may have a similar effect as radial flow on the spectra. The flow value corresponding to the boost received by a deuteron emitted from a completely fused Au+Au system is $\beta\approx0.16$. The actual boost is likely to be considerably less than this, due to pre-equilibrium emission and source expansion.

Thus, it is unlikely that Coulomb repulsion dominates the radial flow signal, especially at higher bombarding energies.

To explore the possibility that directed flow effects are affecting the fits of the spectra, which are integrated over azimuthal angle, we also constructed and fit energy spectra at $\theta_{c.m.}=90°\pm15°$ for particles emitted in the reaction plane ($|\phi_{rp}|<45°$) and out of the reaction plane ($|\phi_{rp}|>45°$), where the reaction plane is determined according to Ref. [21]. Within the uncertainties, no significant difference was observed between the parameters extracted with the $\phi_{rp}$-cut spectra, and the azimuthally-integrated spectra. To increase sensitivity to possible squeeze-out effects [22, 23], the spectra were measured at 90° with respect to the flow axis (as opposed to the beam axis) and cut on $|\phi_{rp}|$. Again no significant difference was observed.

In Figure 2, we plot extracted flow and temperature parameters as a function of bombarding energy for central collisions. Both are seen to increase with bombarding energy. However, if we estimate $E_{thermal}=3 \cdot T/2$ and $E_{flow}=(\gamma-1) \cdot m$, with $\gamma=(1-\beta^2)^{-1/2}$, then our results for central collisions indicate that about 45% of the kinetic energy of deuterons goes into collective radial flow (60% for alphas), with little dependence on bombarding energy. Also shown are the results of fits to energy spectra generated by a QMD model with momentum-dependent interactions [24], and with a BUU transport model that incorporates the emission light mass fragments [5], using a geometric impact parameter distribution in the range b=0-3 fm. These fits are based solely on the shape of the energy spectra, and not on the relative yield. The BUU model only produces fragments up to mass A=3, and the QMD model has been shown to produce too few complex fragments [25]. Furthermore, assumptions underlying the QMD model restrict its validity to bombarding energies below 1 A GeV. Uncertainties in the model fits are of the same order as those for the data. The calculated spectra follow the form of Equation (1) ($\chi^2/\nu\sim1$ for QMD, $\chi^2/\nu\sim2$ for BUU).

Good agreement is observed between the data and QMD values for T and β. Calculations performed with a soft and a hard equation of state (κ=200 and 380 MeV, respectively) yield identical fit parameters, within the uncertainties. The temperature parameters extracted from the BUU spectra (κ=200 and 375 MeV) agree well with the data, but the radial flow values are systematically somewhat high. Little dependence of the parameters on the equation of state is observed for either model.

For non-central collisions and at angles away from 90°, directed flow effects may dominate the energy spectra of emitted particles. An examination of the average energy of emitted particles is illustrative. Neglecting small relativistic effects, the average energy of a particle emitted from a purely thermal source is independent of the particle mass, depending only on the temperature. Superposition of radial flow adds an additional energy component proportional to the particle mass [6]. Figure 3 shows the average kinetic energy of light particles emitted at 90° in the center of mass for the most central Au+Au collisions at E=0.25 and 1.15 A GeV. A roughly linear relationship between $<E_{kin}>$ and A is observed [26]. This linear scaling of $<E_{kin}>$ with A, and not with Z, is further indication that Coulomb effects are not the dominant source of the radial flow signal. Also indicated is the relationship between $<E_{kin}>$ and A expected from the β and T values extracted from the spectral shapes.

Similar linear relationships are observed for particles emitted at forward angles and for all multiplicities above the third multiplicity bin. The slopes of these relationships for Au+Au reactions at E=1.0 A GeV are plotted as a function of multiplicity bin in Fig. 4 for the angular regions $\theta_{cm}$=30°, 60°, 90° ± 15°. The absolute value and multiplicity dependence of the slopes of the relationships at 30°, where directed flow effects are expected to play a more dominant role, differ markedly from those at 60° and 90°. In particular, at 90°, the energy per nucleon induced by collective effects is seen to increase with increasing event centrality, while the mass dependence of the average energy of particles emitted at 30° decreases for the most central collisions, where directed flow is observed to decrease [12]. Indeed, when an additional cut is made such that the particle emitted at 30° is also emitted in the direction opposite of the flow ($p_x<0$, where the positive $p_x$-$p_z$ quadrant contains the major axis of the momentum ellipse), the slope decreases with increasing event centrality, until it is seen to coincide with the slope values at 60° and 90°. Similar cuts have little effect on the slope values at 60° and 90°. Thus, the energy spectra for the most central events, which are expected to be the most spherical in nature, show the clearest signal of "radial" flow. For reference, a geometric impact parameter scale based on the observed charged particle multiplicity [29] is indicated on the top of Figure 4.

In summary, the shapes of midrapidity energy spectra for light particles emitted from Au+Au collisions are well described in terms of a radially-expanding, thermal source. The radial flow value for central collisions is seen to increase as a function of bombarding energy. QMD and BUU model calculations reproduce the temperature and flow parameters satisfactorily, with the BUU exhibiting somewhat too much flow. Very little dependence of the radial flow strength on the equation of state is seen in either model. At $\theta_{cm} \approx 90°$, the collective contribution to the energy is seen to increase with decreasing impact parameter. At forward angles in the flow direction, directed flow is superimposed on the radial flow, while away from the flow direction, the collective energy values converge with those measured at $\theta_{cm} \approx 90°$ for the most central collisions. Radial flow accounts for about 50% of the energy of light particles emitted from central collisions at all bombarding energies.

The authors thank Drs. Georg Peilert and Pawel Danielewicz for the use of their codes and for enlightening discussions. This work was supported in part by the Director, Office of Energy Research, Office of High Energy and Nuclear Physics, Division of Nuclear Physics of the U.S. Department of Energy under Contracts DE-AC03-76SF00098, DE-FG02-89ER40531, DE-FG02-88ER40408, DE-FG02-88ER40412, DE-FG05-88ER40437, and by the National Science Foundation under Grant PHY-9123301.

# Figure Captions

Fig 1) Center-of-mass kinetic energy spectra for light fragments emitted into $\theta_{cm}=90°\pm15°$ from the reaction Au+Au at E = 1.0 A GeV are shown with statistical uncertainties. Fits of the spectra assuming a radially-expanding thermal source (solid lines), and a purely thermal source (dashed lines) are also shown.

Fig 2) Bombarding energy dependence of the temperature and radial flow parameters extracted from the spectra for the most central Au+Au collisions. Fits to spectra generated by a QMD model with soft (dashed lines) and hard (solid lines) equations of state (EOS), and by a BUU model with a soft (dot-dashed lines) and hard (dotted lines) EOS, are also shown; uncertainties for these parameters are on the same order as those for the data.

Fig 3) Average kinetic energy for particles emitted at 90° in the center of mass as a function of the mass number A. Solid lines indicate $<E_{kin}>$ vs A relationships corresponding to $\beta$ and T parameters shown in Fig. 2.

Fig 4) Fitted slopes of the $<E_{kin}>$ vs A relationship as a function of multiplicity bin are shown for $\theta_{cm}=30°,60°,90°\pm 15°$ (filled diamonds, squares, and circles, respectively) for the reaction Au+Au at E=1.0 A GeV. Open diamonds indicate slopes for particles emitted into $\theta_{cm}=30°$ on the negative side of the reaction plane.